\documentclass{appolb}
\usepackage{epsfig}
\begin{document}
\title{Migdal's model and Holography%
\thanks{Presented at The Final Euridice Meeting, 24-27 August 2006, Kazimierz (Poland)}%
}
\author{Oscar Cat\`a\footnote{Fulbright fellow. Formerly at Grup de F{\'\i}sica Te{\`o}rica and IFAE, Universitat
Aut{\`o}noma de Barcelona, 08193 Barcelona, Spain.}
\address{Department of Physics, University of Washington, Seattle WA 98195}
%\and
%the Name(s) of other Author(s)
%\address{and their affiliation}
}
\maketitle
\begin{abstract}
Migdal's model on the spectrum of vector mesons is reassessed. We discuss how its departure from a Pad\'e approximant is closely linked to the issue of quark-hadron duality breakdown. We also show that Migdal's model is not truly a  model of large-$N_c$ QCD.  
\end{abstract}
\PACS{11.15.Tk, 11.25.Tq}
  
\section{Introduction}
More than 30 years ago Migdal proposed a model for hadronic resonances based on rational approximants \cite{Migdal}. The final aim was to determine the spectrum of vector mesons in the large-$N_c$ limit which best fulfills quark-hadron duality. The strategy was to start from the short distance behavior of the vector vacuum polarization function $<VV>$ and build the Pad\'e approximant thereof. Migdal further suggested to modify the continuum limit to better ensure quark-hadron duality. The final result was a spectrum of single poles located at zeroes of the Bessel $J_0$ function.

Migdal's model of resonances was recently revived \cite{Erlich} in the context of the AdS-QCD holographic models. The so called {\it{hard wall model}} \cite{hard} was shown to be the holographic dual of Migdal's model. Interestingly, another holographic model, the so called {\it{soft wall model}} \cite{soft}, was recently identified as the holographic dual of the large-$N_c$ Regge model of Ref. \cite{Shifman}.  

We will reassess Migdal's model with the help of both the 4-dimensional large-$N_c$ Regge model and the holographic duals. We will see that quark-hadron duality breakdown plays a prominent role and look for its geometrical interpretation in holographic models. We will conclude that the modeling of duality breakdown in Migdal's model is not compatible with generic features of large-$N_c$ QCD.   

\section{Migdal's model}
The $[N,M]$ Pad\'e approximant to a function $\Pi_V(q^2)$ is defined as the quotient of two polynomials ${\cal{P}}_M$ and ${\cal{Q}}_N$
\begin{equation}\label{pade}
\Pi_V(q^2)= \Pi_V^{[N,M]}(q^2)+{\cal{R}}_{[N,M]}(q^2)\, , \qquad  \Pi_V^{[N,M]}(q^2)\equiv \frac{{\cal{P}}_M(q^2)}{{\cal{Q}}_N(q^2)},
\end{equation}
such that the first $N+M+1$ derivatives of the function at a point match those of the approximant. ${\cal{R}}_{[N,M]}(q^2)$ is the residue, and convergence of the Pad\'e approximant means that the residue has to vanish as $N,M\rightarrow \infty$. Migdal chose to build the symmetric $[N,N]$ Pad\'e approximant for the parton-model logarithm around a point $q^2=-\mu^2$ in the far Euclidean axis. It is easy to check that the Pad\'e approximant satisfies the following $2N+1$ equations
\begin{equation}\label{padeeq.} 
\frac{d^n}{d(q^2)^n}\bigg[\Pi_V(q^2){\cal{Q}}_N(q^2)-{\cal{P}}_N(q^2)\bigg]\bigg|_{q^2=-\mu^2}=0,\qquad n=0,...,2N \, .
\end{equation}
When $n \geq N+1$, ${\cal{P}}_N(q^2)$ cancels and the denominator ${\cal{Q}}_N(q^2)$ can be determined from the last $N$ equations. Using that $\Pi_V(q^2)$ obeys a once-substracted dispersion relation, one can show that
\begin{equation}\label{pad}
\int_0^{\infty}\frac{dt}{(t+\mu^2)^{n+1}}{\cal{Q}}_N(t)=0,\qquad n=N+1, ... , 2N\, ,
\end{equation} 
which can be shown to lead to Legendre polynomials $P_N^{(0,0)}$,
\begin{equation}\label{result}
{\cal{Q}}_N(q^2)=_{\,\,2}\!\!F_1\left(-N,-N;1;-\frac{q^2}{\mu^2}\right)=(q^2+\mu^2)^N\,P_N^{(0,0)}\left(\frac{\mu^2-q^2}{\mu^2+q^2}\right)\, .
\end{equation}
Eq.(\ref{result}) can now be plugged back in Eq.(\ref{padeeq.}) to yield \cite{Weideman}
\begin{equation}\label{padelog}
\Pi_V^N(q^2)\simeq \frac{2}{(q^2+\mu^2)^N}\,\sum_{k=0}^{N}\left (\begin{array}{c}
k \\
j
\end{array}\right)^2 \left[\frac{H_{N-k}-H_k}{P_N^{(0,0)}(\chi)}\right]\left(-\frac{q^2}{\mu^2}\right)^k\, , \qquad \chi=\frac{\mu^2-q^2}{\mu^2+q^2}\, ,
\end{equation}
which is the Pad\'e approximant to the logarithm, {\it{i.e.}}, in the continuum limit $N\rightarrow \infty$ we recover the logarithm we started from. 
\subsection{Migdal's limit and quark-hadron duality breakdown}
However, in the original approach of Migdal, the continuum limit was taken in a more sophisticated way, namely $q^2<<\mu^2$, $N\rightarrow \infty$ with $N/\mu$ fixed. Under this correlated limit, the Legendre polynomials can be expressed as a Bessel $J_0$ function  
\begin{equation}
\lim_{N\rightarrow \infty} P_N^{(0,b)}\left(1-\frac{\xi^2}{2N^2}\right)=J_0(\xi)\, ,
\end{equation}
and the $<VV>$ Green function takes the form
\begin{equation}\label{migpade}
\Pi_V^N(q^2)=-\frac{4}{3}\frac{N_c}{(4\pi)^2}\left[\log\frac{q^2}{\mu^2}-\pi\frac{Y_0\left(q\Lambda\,\,\right)}{J_0\left(q\Lambda\,\,\right)}\right]\, , \qquad \Lambda=\frac{2N}{\mu}\, .
\end{equation}
It is important to stress that the equation above is no longer the Pad\'e approximant, due to the presence of the Bessel functions. The Pad\'e approximant has the form
\begin{equation}\label{resi}
\Pi_V^N(q^2)=-\frac{4}{3}\frac{N_c}{(4\pi)^2}\log\frac{-q^2}{\mu^2}-\frac{{\cal{R}}_N(q^2)}{{\cal{Q}}_N(q^2)}\, ,
\end{equation}
where the residue is a Meijer G function \cite{Weideman} that vanishes in the continuum limit, thus ensuring convergence. The correlated limits taken by Migdal freeze the residue, thereby spoiling the convergence of the rational approximant. The non-vanishing residue also explains why the resulting spectrum has distinct resonances {\it{after}} $N\rightarrow \infty$.
Migdal's original motivation behind the correlated limits was to enforce quark-hadron duality. However, the correlated limits actually {\it{break}} duality. In the far Euclidean it can be shown that the correlator of Eq.(\ref{migpade}) reduces to ($Q^2=-q^2$)
\begin{equation}\label{inst}
\Pi_V(Q^2)=-\frac{4}{3}\frac{N_c}{(4\pi)^2}\log\frac{Q^2}{\mu^2}+{\cal{O}}(e^{-2Q\Lambda})\, ,\qquad (Q^2\gg 0).
\end{equation}
The last term is not part of the OPE. It is generically referred to as quark-hadron duality violations, defined as  
\begin{equation}\label{analcont}
\Pi_V(q^2)\simeq \Pi_V^{OPE}(q^2)+\Delta(q^2)\, ,\qquad (q^2> 0).
\end{equation}
In Migdal's model, it can be shown that
\begin{equation}
\Delta(q^2)=\frac{N_c}{12\pi}\frac{Y_0\left(q\Lambda\,\,\right)}{J_0\left(q\Lambda\,\,\right)}\, .
\end{equation}
In other words, duality is broken and the duality violating pieces collect the singularities of the spectrum. This is exactly the role that duality violations have to play, since the OPE, being a regular expansion, is unable to capture the resonance poles.
 
There have been some studies devoted to the issue of quark-hadron duality violation \cite{Shifman}. The particular form of Eq.(\ref{inst}) is known to be induced by finite-size singularities. It is important to stress however that this duality breakdown is not what is expected from a large-$N_c$ model of resonances, where infinite-size singularities are present. 

\section{A large-$N_c$ toy model in 4 dimensions}
In the large-$N_c$ limit, the spectral function of the $<VV>$ correlator takes the form 
\begin{equation}\label{ansatz}
\frac{1}{\pi}{\mathrm{Im}}\,\,\Pi_V(t)=\sum_{n=0}^{\infty}F_n^2\delta(t-M_V^2(n))\, .
\end{equation}
We will consider the simplest model in the large-$N_c$ limit with asymptotic Regge behavior, namely \cite{Shifman}
\begin{equation}\label{regge}
F_n^2=F^2,\qquad M_V^2(n)=m_{\rho}^2+an\, .
\end{equation}
Using dispersion relations, the set of Dirac deltas can be resummed to give
\begin{equation}\label{green}
\Pi_V(q^2)=\frac{F^2}{a}\left[\psi\left(\frac{m_{\rho}^2}{a}\right)-\psi\left(\frac{-q^2+m_{\rho}^2}{a}\right)\right]\, .
\end{equation}
The free parameters $F$ and $a$ can be determined by matching to the parton-model logarithm and requiring a vanishing dimension-two condensate
\begin{equation}\label{cond}
c_2=-F^2\left(\frac{m_{\rho}^2}{a}-\frac{1}{2}\right)\, .
\end{equation}
Duality violations can also be determined analytically. Using Eq.(\ref{analcont}),
\begin{equation}\label{duali}
\Delta(q^2)=\frac{N_c}{12\pi} \cot{\bigg[\pi\frac{-q^2+m_{\rho}^2}{a}\bigg]}\, .
\end{equation}
which in the Euclidean decay exponentially as $e^{-|q^2|/\mu^2}$. Notice the difference with Eq.(\ref{inst}). 

\section{Holographic duals}
Both Migdal's model and the toy model introduced in the previous sections have 5-dimensional duals through the AdS-CFT correspondence.\footnote{For a detailed analysis of the relationship between Migdal's model, Pad\'e approximants and holography we refer the reader to \cite{Falk}.} Their action can be written as  
\begin{equation}\label{action}
S=-\int\,d^4x\,dz\,e^{-\Phi(z)}\sqrt{g}\frac{1}{4g_5^2}{\mathrm{Tr}}\,\bigg[(F_{{\hat{\mu}}{\hat{\nu}}}F^{{\hat{\mu}}{\hat{\nu}}})_L+(F_{{\hat{\mu}}{\hat{\nu}}}F^{{\hat{\mu}}{\hat{\nu}}})_R\bigg]\, ,
\end{equation}
where $g$ is the AdS metric
\begin{equation}
g_{\hat{\mu}\hat{\nu}}\,dx^{\hat{\mu}}dx^{\hat{\nu}}=\frac{1}{z^2}(\eta_{\mu\nu}dx^{\mu}dx^{\nu}+dz^2)\, ,\qquad \eta_{\mu\nu}={\mathrm{diag}}(-1,1,1,1)\, ,
\end{equation}
and $\Phi(z)$ is the dilaton field. In the {\it{hard wall model}}, $\Phi(z)=\phi,\,\epsilon\leq z\leq \Lambda$, where $z=\epsilon$ and $z=\Lambda$ are the four-dimensional boundary branes. In this model, the solution to $<VV>$ is given by Eq.(\ref{migpade}). Holography therefore provides a nice geometrical interpretation of the infrared scale $\Lambda$ of Migdal's model.

The so called {\it{soft wall model}} was originally proposed as a AdS-QCD model with built-in Regge behavior. The theory does not have an infrared brane, and instead the dilaton field has a quadratic profile, $\Phi(z)\sim cz^2$. The {\it{soft wall model}} is the holographic dual to our toy model with $m_{\rho}^2=4c=a$ \cite{soft}.
Notice that modifications on the infrared of 5-dimensional holographic models correspond to different spectrums and therefore to different violations of quark-hadron duality. In the language of holography a realistic large-$N_c$ spectrum is achieved, as illustrated in the {\it{soft wall model}}, by the removal of the infrared brane.  
\subsection{Some phenomenology of the soft-wall model}
The {\it{soft wall model}} has two independent parameters, $g_5$ and $c$. $c$ is determined by $m_{\rho}$ and $g_5$ can be fixed by imposing matching of the model to the perturbation theory logarithm. Thus,
\begin{equation}
g_5^2=\frac{12\pi^2}{N_c}\, , \qquad c=\frac{m_{\rho}^2}{4}\simeq 150\,{\mathrm{MeV}}^2\, .
\end{equation}
By matching to our toy model one immediately finds that duality violations take the form \cite{Cata}
\begin{equation}
\Delta(q^2)=\frac{\pi}{2g_5^2}\cot{\left[\pi\frac{-q^2+4c}{4c}\right]}\, .
\end{equation}
Unfortunately, $m_{\rho}^2=a$ predicts a Regge slope $a$ and decay constants $F$ at about half their experimental values. Furthermore, using Eq.(\ref{cond}), the {\it{soft wall model}} yields a negative dimension-two condensate.
Interestingly, $2 m_{\rho}^2=a$ would bring both $a$ and $F$ closer to experimental values as well as making the dimension-two condensate vanish.
\section{Discussion}
We have shown that Migdal's model for the spectrum of vector mesons in QCD is {\it{not}} a Pad\'e approximant. Pad\'e approximants are discretizations of functions and, if convergence is to be fulfilled, in the continuum limit $N\rightarrow \infty$ they should yield the function one has started with. In Migdal's model, one starts with the parton model logarithm and ends in a theory of discrete resonances, even after $N\rightarrow \infty$.

This departure from the Pad\'e approximant is due to the modified continuum limit, which also generates an infrared scale $\Lambda$. This infrared scale has a nice geometrical interpretation as the infrared boundary brane of 5-dimensional holographic models. The scale breaks the convergence of the Pad\'e approximant and is actually a modeling of quark-hadron duality breakdown. The existence of duality breakdown is a necessary ingredient in any model of resonances: from the operator product expansion one cannot reach the spectrum. Notice however that Migdal's modeling of duality breakdown is based on a manipulation of the asymptotic behavior of $<VV>$. Therefore it is {\it{by construction}} unable to distinguish models with the same asymptotics.

Migdal's spectrum (and hence the spectrum in the {\it{hard wall model}}) tends to the Pad\'e approximant at large spacelike momenta and hence reproduces the perturbative logarithm. However, on the physical axis, due to the presence of duality breakdown, the picture is very different: the Pad\'e tries to reconstruct the logarithmic branch cut by piling simple poles arbitrarily close, thus invalidating any possible interpretation of them as resonances in a large-$N_c$ model. Inclusion of duality breakdown is thus, contrary to what Migdal claimed, essential.

\section*{Acknowledgements}
The author is supported by the Fulbright Program and the Spanish Ministry of Education and Science under grant no. FU2005-0791.
  
\end{document}